\documentclass[aps,prb,amsmath,showpacs,preprint]{revtex4}
\usepackage{bm}
\usepackage{amsmath}
\usepackage{amssymb}
\usepackage{amsfonts}
\usepackage{graphicx}
\usepackage{color}

\begin{document}

\title{ The Induced Charge Generated By The Potential Well In Graphene}

\author{Alexander I. Milstein and Ivan S. Terekhov}
\affiliation{Budker Institute of Nuclear Physics, 630090
Novosibirsk, Russia}


\begin{abstract}
The induced charge density, $\rho_{ind}(\bm r)$, generated in
graphene by the potential well of the finite radius $R$  is
considered.  The result for $\rho_{ind}(\bm r)$ is derived for large
distances $r\gg R$. We also obtained the induced charges outside of
the radius $r\gg R$ and inside of this radius for subcritical and
supercritical regimes. The consideration is based on the convenient
representation of the induced charge density via the Green's
function of electron in the field.

\end{abstract}


\maketitle

\section{Introduction}

As known, the induced charge density, $\rho_{ind}({\bm r})$, in the
external electric field appears due to vacuum polarization. In the
field  of heavy nucleus, this important effect of Quantum
Electrodynamics (QED) was investigated in detail in many papers,
see, e.g., Refs. \cite{WichmannKroll,McLerran,Mil2,Zeld}. New
possibilities to study vacuum polarization in QED at large coupling
constant  have appeared after recent successful fabrication of a
monolayer graphite (graphene), see Ref. \cite{Novoselov1} and recent
Review \cite{CN09}. The single electron dynamics in  graphene is
described by a massless two-component Dirac equation
\cite{Wallece,McClure,Semenoff84,Gonzalez} so that graphene
represents a  two-dimensional (2D) version of massless QED. On the
one hand, this version is essentially simpler than conventional QED
because effects of retardation are absent due to  instant Coulomb
interaction between electrons. On the other hand, the  ``fine
structure constant'' $\alpha=e^2/\hbar v_F$ is of order of unity
since the Fermi velocity $v_F \approx 10^6 \mbox{m/s} \approx c/300$
($c$ is the velocity of light), and therefore we have  a
strong-coupling version of QED. Below we set $\hbar=c=1$.

Screening of charged impurity  in graphene can also be treated in
terms of vacuum polarization
\cite{DiVincenzo,nomura,Ando,Hwang,Katsnelson,Shytov,
pereira,Subir,Fogler,tmks07,Kotov08,Kotov08a}. Investigation of
impurity screening is important for  understanding of the dependence
of transport properties  on the impurity concentration. There are
two regimes for the Coulomb impurity in the gapless graphene,
subcritical and supercritical. In the subcritical regime, it is
shown in the leading order in $\alpha$ and exactly in the Coulomb
potential that  the induced charge is localized at the impurity
position, see \cite{Shytov,Subir,pereira,tmks07,Kotov08}. In the
supercritical regime, vacuum polarization in the Coulomb field has
been recently considered  in Refs.\cite{Katsnelson,Kotov08a}. In
this case, the induced charge density is not localized   at the
impurity position due to the effect similar  to that of $e^+e^-$
pair production in 3D QED in the electric field of supercritical
heavy nuclei. In the present paper, we answer to the question
whether the phenomenon of the induced charge localization also exist
in the potential well of finite size $R$ and depth $U_0$. Namely, we
calculate the asymptotics  of $\rho_{ind}({\bm r})$ in the field of
an
 azimuthally  symmetric potential well at large distances $r\gg R$. We
apply  the method  suggested in Ref.~\cite{LeeMil} for calculation
of  the finite nuclear size effect  on the induced charge density at
large distances in a strong Coulomb field  in 3D QED. We show that
there are also subcritical and supercritical regimes in this
problem. However, the induced charge is not localized at $r\lesssim
R$  in the subcritical regime and has  power "tail" in contrast to
the case of the Coulomb field. In the vicinity of transition from
the subcritical regime to the  supercritical one,  small variation
of the potential parameters drastically changes the induced charge
density. We demonstrate that this fact is not related to  the
smoothness of the potential well. The attempt to calculate the
induced charge distribution in the potential well in graphene was
previously performed in Ref.\cite{DiVincenzo}. The authors of this
paper used the method which akin to that used at calculation of
conventional Friedel oscillations. However, our results for the
induced charge density differ from that obtained in
Ref.\cite{DiVincenzo} mainly due to the mistake performed in
Ref.\cite{DiVincenzo} at the calculation of the phase shift.

The paper is organized as follows. In Section \ref{section2} we
derive the general expression for the induced charge density
convenient for calculation of the asymptotics at large distances. In
Section \ref{section3} we consider  the Green's function of electron
in an azimuthally    symmetric potential and   use this function in
calculations of $\rho_{ind}({\bm r})$ in Section \ref{section4}.
Critical values of $g$ are discussed in Section \ref{section5}
calculating the  scattering phase shifts of electron wave function
in the field of the potential well. The induced charges outside of
the radius $r\gg R$ and inside of this radius for subcritical and
supercritical regimes are considered in Section \ref{section6}.
Finally, in Section \ref{section7} the main conclusions of the paper
are presented.

\section{General discussion}\label{section2}
  In graphene, the induced charge density  in the potential $U(r)$ have the form
\begin{eqnarray}\label{Inducedrho}
\rho_{ind}({\bm r})=-ieN\int_{C}\frac{d\epsilon}{2\pi}\mathrm{Tr}\{
G({\bm r},{\bm r}|\epsilon)\}\,,
\end{eqnarray}
where  $N=4$ reflects the spin and valley degeneracies, and the
 Green's function $G({\bm r},{\bm r}'|\epsilon)$
satisfies the equation
\begin{equation}
\label{gfe} \left[\epsilon-U(r) -v_F\bm\sigma\cdot\bm p
\right]G({\bm r},{\bm r}'|\epsilon)=\delta(\bm r-{\bm r}')I.
\end{equation}
Here ${\bm{\sigma}}=(\sigma_1,\sigma_2)$, and $\sigma_i$ are the
Pauli matrices; ${\bm p}=(p_x,p_y)$ is the momentum operator, $\bm
r=(x,y)$, and $I=\mathrm{diag}\{1,1\}$. The matrixes ${\bm{\sigma}}$
act on the pseudo-spin variables and the spin degrees of freedom are
taken into account in a factor $N$. According to the Feynman rules,
the contour of integration over $\epsilon$ goes below the real axis
in the left half-plane and above the real axis in the right
half-plane of the complex $\epsilon$ plane. Using the analytical
properties of the Green's function, we deform the contour of
integration with respect to $\epsilon$ so that it coincides finally
with the imaginary axis. Then we follow Ref.\cite{LeeMil} and write
the equation for the Green's function in the form
\begin{eqnarray}\label{GreenFunctionExp}
G({\bm r},{\bm r}'|i\epsilon) =G^{(0)}({\bm r},{\bm r}'|i\epsilon)
+\int d{\bm r}_1 d{\bm r}_2G^{(0)}({\bm r},{\bm
r}_1|i\epsilon)\left[U({\bm r}_1)\delta({\bm r}_1-{\bm r}_2)\right.
\nonumber\\
  +\left.U({\bm r}_1)G({\bm r}_1,{\bm
r}_2|i\epsilon)U({\bm r}_2)\right]G^{(0)}({\bm r}_2,{\bm
r}'|i\epsilon)\, ,
\end{eqnarray}
where $G^{(0)}({\bm r},{\bm r}'|i\epsilon)$ is the solution of
Eq.(\ref{gfe}) at zero  external field.

It is convenient to represent $\rho_{ind}(r)$ as a sum
\begin{eqnarray}\label{InducedCharge}
\rho_{ind}(r) = \rho_{ind}^{(1)}(r)+\rho_{ind}^{(2)}(r)\, ,
\end{eqnarray}
where $\rho_{ind}^{(1)}(r)$ is the linear in $U(r)$ contribution and
$\rho_{ind}^{(2)}(r)$ is the contribution of high order in $U(r)$
terms. It follows from Eqs.(\ref{Inducedrho}) and
(\ref{GreenFunctionExp}) that
\begin{eqnarray}\label{InducedCharge1}
\rho_{ind}^{(1)}(r)
=eN\int_{-\infty}^{\infty}\frac{d\epsilon}{2\pi}\int d{\bm r}_1
 \mathrm{Tr}\Bigl\{G^{(0)}({\bm r},{\bm
r}_1|i\epsilon)U( r_1)G^{(0)}({\bm r}_1,{\bm
r}|i\epsilon)\Bigr\}\, ,
\end{eqnarray}
\begin{eqnarray}\label{InducedCharge2}
\rho_{ind}^{(2)}(r)
=eN\int_{-\infty}^{\infty}\frac{d\epsilon}{2\pi}\int d{\bm r}_1
d{\bm r}_2 \mathrm{Tr}\Bigl\{G^{(0)}({\bm r},{\bm
r}_1|i\epsilon)U({r}_1)G({\bm r}_1,{\bm r}_2|i\epsilon)U({
r}_2)G^{(0)}({\bm r}_2,{\bm r}|i\epsilon)\Bigr\}\, .
\end{eqnarray}
Formulas (\ref{InducedCharge1}) and (\ref{InducedCharge2}) are very
convenient for calculation of the induced charge density at large
distances.

\section{Green's function in an  azimuthally    symmetric
potential}\label{section3}

The free Green's function $G^{(0)}({\bm r},{\bm r}'|i\epsilon)$ is
given by
\begin{eqnarray}\label{GreenFunction0}
G^{(0)}({\bm r},{\bm r}'|i\epsilon)= -\frac{i\epsilon}{2\pi}
\left[K_0(|\epsilon|\xi)-\mathrm{sign}(\epsilon)\frac{(\bm\sigma\cdot{\bm\xi
)}}{\xi}K_1(|\epsilon|\xi)\right]\, ,
\end{eqnarray}
 where $\bm\xi={\bm r}-{\bm r}'$, and $K_{0,1}(x)$ are the modified Bessel functions of the third kind.
Let us  represent the electron Green's function $G(\bm r,{\bm
r}'|\epsilon)$ in an
 azimuthally  symmetric potential $U(r)$ in the form
\begin{eqnarray}
\label{GreenFunctionGeneral} G({\bm r},{\bm r}'|\epsilon)=
\frac{1}{2\pi}\sum_{m=-\infty}^{\infty} e^{im(\phi-\phi')}
\begin{pmatrix}
A_m(r,r'|\epsilon) & -ie^{-i\phi'}B_m(r,r'|\epsilon)\\
ie^{i\phi}C_m(r,r'|\epsilon)&e^{i(\phi-\phi')}D_m(r,r'|\epsilon)
\end{pmatrix}\,,
\end{eqnarray}
and use the relation
\begin{eqnarray}
\delta({\bm r}-{\bm r}')=
\frac{\delta(r-r')}{2\pi\sqrt{rr'}}\sum_{m=-\infty}^{\infty}
e^{im(\phi-\phi')} \, .
\end{eqnarray}
Then, from  Eq.(\ref{gfe})  we obtain the  equations
\begin{eqnarray}\label{system}
&&(\epsilon-U(r))A_m-\dfrac{\partial C_m}{\partial r}
-\dfrac{m+1}{r}C_m=\dfrac{\delta(r-r')}{\sqrt{rr'}}\, ,\nonumber \\
&&(\epsilon-U(r))C_m+\dfrac{\partial A_m}{\partial
r}-\dfrac{m}{r}A_m=0\, ,
\end{eqnarray}
and the relations  $D_m= A_{-m-1}$ and $B_m=- C_{-m-1}$. Therefore,
to find the Green's function in a azimuthally  symmetric potential,
it is sufficiently to solve  equations (\ref{system}).

\section{An induced charge density  at large
distances}\label{section4}

To calculate the asymptotics  of the function $\rho_{ind}^{(1)}$ at
distances $r\gg R$, where $R$ is a typical size of the potential, we
can put $r_1=0$ in the arguments of the free Green's functions in
Eq.(\ref{InducedCharge1}). After that we take the integral over
$\epsilon$ and obtain:
\begin{eqnarray}\label{InducedCharge1Asymptotic}
\rho_{ind}^{(1)}(r) =\frac{eN}{16\, r^3}\int d{ r}'\,r'U( r')\, .
\end{eqnarray}
One can see that the induced charge density in the leading order in
the external field goes to zero at large distances as $1/r^3$.

Let us consider  the function $\rho_{ind}^{(2)}(r)$ at $r\gg R$. We
substitute Eqs.(\ref{GreenFunctionGeneral}) and
(\ref{GreenFunction0}) to  Eq.(\ref{InducedCharge2}), put  $r_1=0$
and $r_2=0$ in the arguments of the free Green's function, and take
the integral over angels of the vectors ${\bm r}_1$ and ${\bm r}_2$.
Then we  obtain
\begin{eqnarray}\label{InducedCharge2Asymptotic}
\rho_{ind}^{(2)}(r)
=-\frac{eN}{2\pi^2}\int\limits_{-\infty}^{\infty}d\epsilon\,\epsilon^2
\left[K_0^2(|\epsilon|r)-K_1^2(|\epsilon|r)\right]
\int\limits_0^\infty\!\!\int\limits_0^\infty
d{r}_1 d{r}_2 \, r_1 r_2
U({r}_1)U({r}_2)A_0({r}_1,{r}_2|i\epsilon)\, .
\end{eqnarray}
Here $A_0({r}_1,{r}_2|i\epsilon)$ is the solution of
Eq.(\ref{system}) at $m=0$. Note  that
Eq.(\ref{InducedCharge2Asymptotic}) includes the contributions of
the terms with  $m=0$ and $m=-1$ in the Green's function
(\ref{GreenFunctionGeneral}), since $D_{-1}=A_0$. It is convenient
to introduce the functions
\begin{eqnarray}\label{functions}
a(r,\epsilon) = \int_0^\infty dr'r'U(r')A_0(r,r'|i\epsilon)\, ,
\quad c(r,\epsilon) = \int_0^\infty dr'r'U(r')C_0(r,r'|i\epsilon) \,
,
\end{eqnarray}
Let us  multiply both sides of the equations (\ref{system})
 by $r'U(r')$, and then take the integral over $r'$ from zero to
infinity. As a result we obtain the following equations for the
functions $a(r,i\epsilon)$ and $c(r,i\epsilon)$:
\begin{eqnarray}\label{system2}
&&(i\epsilon-U(r))a(r,\epsilon)-\dfrac{\partial c(r,\epsilon)}
{\partial r}-\dfrac{c(r,\epsilon)}{r}=U(r)\, ,\nonumber \\
&&(i\epsilon-U(r))c(r,\epsilon)+\dfrac{\partial
a(r,\epsilon)}{\partial r}=0\, .
\end{eqnarray}
The boundary conditions for these equations are
$a(0,\epsilon),c(0,\epsilon)<\infty$, and $\lim\limits_{r\to\infty}
a(r,\epsilon)=\lim\limits_{r\to\infty}c(r,\epsilon)=0$. In terms of
the function  $a(r,i\epsilon)$, Eq.(\ref{InducedCharge2Asymptotic})
has the form
\begin{eqnarray}\label{InducedCharge2Asymptotic2}
\rho_{ind}^{(2)}(r)
=-\frac{eN}{2\pi^2}\int_{-\infty}^{\infty}d\epsilon\,\epsilon^2
\left[K_0^2(|\epsilon|r)-K_1^2(|\epsilon|r)\right]\int_0^\infty dr'
r' U(r') a(r',\epsilon)\, .
\end{eqnarray}
Then we pass in this equation from the variable $\epsilon$ to the
variable $E=r\epsilon$ and replace  $a(r',E/r)$ on $a(r',0)$ at
$r\gg R$. We can do that because the integral over $E$ converges at
$E\sim 1$ due to the properties of the $K$- functions. After this
replacement we take the integral  over $E$ and arrive at the
following expression for the asymptotics  of $\rho_{ind}^{(2)}(r)$:
\begin{eqnarray}\label{Rho2MainAsymptotic}
\rho_{ind}^{(2)}(r) =\frac{eN}{16\, r^3}\int_0^\infty dr' r'
U(r')a(r',0)\, .
\end{eqnarray}
Thus, the function $\rho_{ind}^{(2)}(r)$ has the same behavior at
large distances as $\rho_{ind}^{(1)}(r)$.

Let us consider a simple example of the  potential,
$U(r)=-U_0\theta(R-r)$, where $\theta(x)$ is the step function, $R$
is the radius of the potential well. The solution $a(r,0)$ of Eq.
(\ref{system2}) is
\begin{eqnarray}\label{solution_a0}
a(r,0)=\left\{
\begin{array}{ll}
\frac{J_0(U_0 r)}{J_0(U_0 R)}-1 \, ,& r<R \, ,\\
0\, ,& r>R\, , \\
\end{array}\right.
\end{eqnarray}
where $J_n(x)$ is the Bessel function. Using this solution,  we find
the sum of the contributions Eq.(\ref{InducedCharge1Asymptotic}) and
Eq.(\ref{Rho2MainAsymptotic}),
\begin{eqnarray}\label{rho1}
\rho_{ind}(r) =-\frac{eN  J_1(g) R}{16 J_0(g)r^3}\, ,
\end{eqnarray}
where $g=U_0 R$ is the effective dimentionless coupling  constant.
The induced charge density (\ref{rho1}) is the odd function of the
parameter $g$, which corresponds to the Furry theorem in QED. The
formula (\ref{rho1}) contains singularities at the critical values
of $g=g_c$ satisfying  the equation $J_0(g_c)=0$. In our case, the
first three values are $g_c\thickapprox 2.41,\, 5.52,\,8.65$.
Existence of such singularities is not related to  strong variation
of our potential around  the point $r=R$. We found numerically the
first three critical values of $g=U_0R$  for the smooth potentials
$U_1(r)=-U_0e^{-r/R}$  and $U_2(r)=-U_0e^{-r^2/R^2}$. In the first
case,  $g_c\thickapprox 2.87,\, 5.9,\, 9.0$. In the second case,
$g_c\thickapprox 2.7,\, 5.7,\,8.0$. We see that the corresponding
numerical values of $g_c$  are close to each other.

Actually,  singularities in  Eq.(\ref{rho1}) have appeared as a
result of substitution  $a(r',E/r)\rightarrow a(r',0)$ in
Eq.(\ref{Rho2MainAsymptotic}), which is not valid in the vicinity of
$g=g_c$ since Eq.(\ref{solution_a0}) has no sense at $g=g_c$. In the
vicinity of $g=g_c$, it is necessary to perform calculation of the
integrals in Eq.(\ref{InducedCharge2Asymptotic2}) more accurately.
For the step-like potential $U(r)=-U_0\theta(R-r)$, the solution of
Eq. (\ref{system2}) at $\epsilon\ne 0$ has the form
\begin{eqnarray}\label{solution_a}
a(r,\epsilon)=\left\{
\begin{array}{ll}
\frac{1}{\gamma} J_0((U_0+i\epsilon)r)-U_0/(U_0+i\epsilon)\, ,& r<R \, ,\\
\beta K_0(|\epsilon|r)\, ,& r>R\, , \\
\end{array}\right.
\end{eqnarray}
\begin{eqnarray}\label{solution_c}
c(r,\epsilon)=\left\{
\begin{array}{ll}
\frac{1}{\gamma} J_1((U_0+i\epsilon)r)\, ,& r<R \, ,\\
-i\beta \mathrm{sign}(\epsilon) K_1(|\epsilon|r)\, ,& r>R\, .\\
\end{array}\right.
\end{eqnarray}
Taking into account  continuity of the functions $a(r,\epsilon)$ and
$c(r,\epsilon)$ at $r=R$, we obtain
\begin{eqnarray}\label{solution_alpha}
\gamma=\left(1+\frac{i\epsilon}{U_0}\right)\left[J_0((U_0+i\epsilon)R)-i\,
\mathrm{sign}(\epsilon)
J_1((U_0+i\epsilon)R)\frac{K_0(|\epsilon|R)}{K_1(|\epsilon|R)}\right]\,
.
\end{eqnarray}
Then we  substitute  Eqs.(\ref{solution_a}) and
(\ref{solution_alpha}) to  Eq.(\ref{InducedCharge2Asymptotic2}) and
take the  integral over $r'$. As above,  the main contribution to
the integral over $\epsilon$ at $r\gg R$ is given by the region
$\epsilon\lesssim 1/r$, so that we can use the relations
  $\epsilon R\ll 1$ and $\epsilon/U_0\ll 1$. Finally  we find
the expression for the sum of $\rho_{ind}^{(1)}(r)$ and
$\rho_{ind}^{(2)}(r)$ at large distances,
\begin{eqnarray}\label{rho}
\rho_{ind}(r) =\frac{eN J_0(g) J_1(g)
R}{\pi^2}\int_{0}^{\infty}d\epsilon\,\epsilon^2
\frac{K_0^2(\epsilon r)-K_1^2(\epsilon
r)}{J^2_0(g)+J^2_1(g)(\epsilon R)^2\ln^2(\epsilon R)} \, .
\end{eqnarray}
This expression is valid at arbitrary value of the coupling constant
$g=U_0 R$. If $|J_0(g)|\gg (R/r)\ln(r/R)$, then it is possible to
neglect the second term in the denominator of the integrand, and we
return to the expression (\ref{rho1}). If $g$ is close to some $g_c$
so that $|J_0(g)|\ll (R/r)\ln(r/R)\ll 1$, we obtain
\begin{eqnarray}\label{rhosmallDisct}
\rho_{1}(r) =-\frac{eN  \mathrm{sign}(g-g_c)}{2 \pi\,r^2
\ln|g-g_c|}\, .
\end{eqnarray}
In this case  the induced charge density diminishes as  $1/r^2$ and
has opposite sign for $g<g_c$ and $g>g_c$. In terms of distances,
the asymptotics (\ref{rhosmallDisct}) is valid at
$$1\ll r/R\ll -\ln|g-g_c|/|g-g_c|\,.$$
 At  $r/R\gg-\ln|g-g_c|/|g-g_c|$ and $|g-g_c|\ll 1$, we have
 (see Eq.(\ref{rho1}))
\begin{eqnarray}\label{rho11}
\rho_{2}(r) =\frac{eN  R}{16 r^3(g-g_c)}\, ,
\end{eqnarray}

\begin{figure}[h]
\includegraphics[scale=1.]{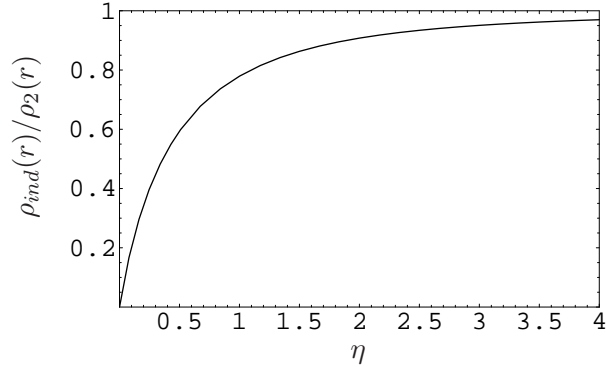}
\begin{picture}(0,0)(0,0)
\put(-102,-8){$\eta$}
 \put(-230,45){\rotatebox{90}{$\rho_{ind}(r)/\rho_{2}(r)$}}
 \end{picture}
\caption{Ratio  $\rho_{ind}(r)/\rho_{2}(r)$ at $r\gg R$ and
$|g-g_c|\ll1$ as a function of $\eta=-r |g-g_c|/(R\ln|g-g_C|)$. The
asymptotics $\rho_2(r)$ is  given by Eq.(\ref{rho11}) and
$\rho_{ind}$ by Eq.(\ref{rho}).}\label{AS}
\end{figure}
In order to illustrate the transition from the asymptotics
(\ref{rhosmallDisct}) to the asymptotics (\ref{rho11}), we consider
the ratio $\rho_{ind}(r)/\rho_{2}(r)$ at  $r\gg R$ and
$|g-g_c|\ll1$. In this case this ration depends only on the variable
$\eta=-r |g-g_c|/(R\ln|g-g_C|)$. The dependence of
$\rho_{ind}(r)/\rho_{2}(r)$ on $\eta$ is shown in Fig.\ref{AS}. We
see that $\rho_{ind}(r)\approx\rho_{2}(r)$ already at $\eta\simeq
2$.

\section{Critical values of $g$ and scattering problem}\label{section5}
It is possible to explain critical values of $g$ using the approach
based on the  scattering problem, as it is usually performed at the
consideration of   Friedel oscillations, see Ref.\cite{Kittel}.
Writing the wave function of electron as
\begin{eqnarray}
\label{psi} \psi({\bm r})=
\begin{pmatrix}
u_m(r)e^{im\phi}\\
id_m(r)e^{i(m+1)\phi}
\end{pmatrix}\,,
\end{eqnarray}
we obtain  equations for the functions $u_{m}(r)$ and $d_{m}(r)$,
cf. Eq.(\ref{system}),
\begin{eqnarray}\label{systempsi}
&&(\epsilon-U(r))u_m-\dfrac{\partial d_m}{\partial r}
-\dfrac{m+1}{r}d_m=0\, ,\nonumber \\
&&(\epsilon-U(r))d_m+\dfrac{\partial u_m}{\partial
r}-\dfrac{m}{r}u_m=0\, .
\end{eqnarray}

The solution of this equations in the the step-like potential has
the form (common normalization factor is omitted):
\begin{eqnarray}\label{solution_u}
u_m(r)=\left\{
\begin{array}{ll}
J_m(|U_0+\epsilon|r)\, ,& r<R \, ,\\
\mu_m J_m(|\epsilon|r)+\nu_m N_m(|\epsilon|r)\, ,& r>R\, , \\
\end{array}\right.
\end{eqnarray}
\begin{eqnarray}\label{solution_d}
d_m(r)=\left\{
\begin{array}{ll}
 \mbox{sign}(U_0+\epsilon)\,J_{m+1}(|U_0+\epsilon|r)\, ,& r<R \, ,\\
\mbox{sign}(\epsilon)\,[\mu_m J_{m+1}(|\epsilon|r)+\nu_m N_{m+1}(|\epsilon|r)]\, ,& r>R\, .\\
\end{array}\right.
\end{eqnarray}
Here $N_{m}(x)$ are the  Bessel functions of the second kind. From
continuity of the functions $u_m(r)$ and $d_m(r)$ at $r=R$, we have
\begin{eqnarray}\label{solutionmunu}
\mu_m&=&-\frac{\pi|\epsilon|R}{2}\Big[J_m(|U_0+\epsilon|R)N_{m+1}(|\epsilon|R)\nonumber\\
&&-J_{m+1}(|U_0+\epsilon|R)N_{m}(|\epsilon|R)\mathrm{sign}(\epsilon)
\mathrm{sign}(\epsilon+U_0)\Big]\, ,\nonumber\\
\nu_m&=&-\frac{\pi|\epsilon|R}{2}\Big[-J_m(|U_0+\epsilon|R)J_{m+1}(|\epsilon|R)\nonumber\\
&&+J_{m+1}(|U_0+\epsilon|R)
J_{m}(|\epsilon|R)\mathrm{sign}(\epsilon)
\mathrm{sign}(\epsilon+U_0)\Big]\, .
\end{eqnarray}
Using the asymtotics of the Bessel functions at large value of
argument, we find  the phase shift
$\delta_m(\epsilon)=-\arctan(\nu_m/\mu_m)$. Critical values of $g$
are given by the solution of the equation $J_m(g_c)=0$ at $m\ge 0$
and $J_{|m|-1}(g_c)=0$ at $m<0$. Taking into account the asymptotics
$$N_0(x)\approx \frac{2\ln x}{\pi}\quad ,\quad N_{|m|}(x)\approx
-\frac{2^{|m|}(|m|-1)!}{x^{|m|}\pi}$$ at $x\ll 1$, we find for
$|\epsilon|R\ll 1$, $|\epsilon|\ll U_0$, and $g=U_0R$  close to
$g_c$
\begin{eqnarray}\label{deltam0}
\delta_0(\epsilon)&=&\arctan\left[ \frac{{\pi\over 2} \epsilon R}
{\epsilon R\ln(|\epsilon|R)-(g-g_c)}\right]
\, ,\nonumber\\
\delta_{m}(\epsilon)&=&-\arctan\left[\frac{\pi(\epsilon R)^{2m+1}}
{2^{2m}m!(m-1)!\,[(2m+1)\,\epsilon R+2m(g-g_c)]}\right] \,
\mbox{at}\, m>0\, ,
\end{eqnarray}
 and $\delta_{-|m|}(\epsilon)=\delta_{|m|-1}(\epsilon)$.
  If  $\epsilon <0$, which corresponds to electrons inside Fermi surface, and $g<g_c$, then
 $\delta_m$ is always  small. For $g>g_c$,  the phase shift $\delta_m(\epsilon)$ can be
 equal to $\pm\pi/2$ at some $\epsilon <0$. That means the
 appearance at  $g=g_c$ of the additional quasi-bound state on the Fermi
 surface.

 Calculation of the phase  shift in the step-like potential was previously
  performed in Ref.\cite{DiVincenzo}.
However, the coefficients corresponding to $\mu_m$ and $\nu_m$,
Eq.(\ref{solutionmunu}), were found in Ref.\cite{DiVincenzo} by
matching  the function $u_m(r)$ and its first derivative at $r=R$,
instead of matching  the functions $u_m(r)$ and $d_m(r)$. It is easy
to check that the first derivative of $u_m(r)$ is not a continuous
function in the point $r=R$. As a consequence, the asymptotics of
the induced charge density at large distances obtained in
Ref.\cite{DiVincenzo} is not correct.

\section{An induced charge }\label{section6}

Let us consider the induced charge $Q_>(r)$ outside  of the radius
$r\gg R$,
\begin{eqnarray}\label{Charge}
Q_>(r)&=&2\pi\int_r^\infty dr'r'\rho_{ing}(r')\nonumber\\
&=&-\frac{eN J_0(g) J_1(g) Rr^2}{\pi}
\int_{0}^{\infty}d\epsilon\,\epsilon^2 \frac{K_0^2(\epsilon
r)+K_0(\epsilon r)K_2(\epsilon r)-2K_1^2(\epsilon
r)}{J^2_0(g)+J^2_1(g)(\epsilon R)^2\ln^2(\epsilon R)} \, .
\end{eqnarray}
For $|J_0(g)|\gg (R/r)\ln(r/R)$, we have
\begin{eqnarray}\label{ChargeA}
Q_>(r)=-\frac{eN\pi RJ_1(g)}{8J_0(g)\,r} .
\end{eqnarray}
In the case $|g-g_c|\ll (R/r)\ln(r/R)$, we find with logarithmic
accuracy
\begin{eqnarray}\label{ChargeB}
Q_>(r) =eN
\mathrm{sign}(g-g_c)\left(1+\frac{\ln(r/R)}{\ln|g-g_c|}\right)\, .
\end{eqnarray}
Since $N=4$, then  $Q_>(r)/e$ tends to the integer number $
N\mathrm{sign}(g-g_c)$ at $g\rightarrow g_c$, having opposite sign
for $g<g_c$ and $g>g_c$.

Let us discuss  the  induced charge $Q_<(r)$ inside  of the radius
$r\gg R$.  Since the total induced charge $Q_{tot}=Q_>(r)+Q_<(r)$ is
zero for the potential well at $g$  less than the minimal $g_c$, we
have $Q_<(r)=-Q_>(r)$ for such value of $g$. Note that $Q_{tot}$ is
not zero for massless electron in graphene in the Coulomb potential
$U_C(r)=-Z\alpha/r$ even in the subcritical regime $Z\alpha<1/2$,
see Ref.\cite{tmks07},  due to zero mass of a particle and slow
decreasing of a Coulomb potential at large distances. For $g$ larger
than the minimal $g_c$, the total induced charge is already not
equal to zero due  to the effect similar to $e^+e^-$ pair production
in the electric field of superheavy nucleus \cite{Zeld,Greiner
Muller Rafelski}. In this case $Q_{tot}=eNM$, where $M$ is a number
of $g_c$ less than $g$, so that $Q_<(r)=-Q_>(r)+eNM$. The quantity
$M$ is nothing but  the number of the quasi-bound states at a given
value of $g$, see discussion in Section\ref{section5}. The explicit
values of $g_c$ are given by zeros of the Bessel functions, as it is
pointed out in Section\ref{section5}.

In Section \ref{section4} and this section we have considered the
contributions of the angular momenta $m=0$ and $m=-1$ in the Green's
function (\ref{GreenFunctionGeneral}) to induced charge density and
$Q_>(r)$ at large distances. Of course, the contributions of $m>0$
and $m<-1$ are not zero though they are strongly suppressed by some
power of $R/r$ even in the vicinity of the corresponding critical
points. However,  $M$ in $Q_{tot}=eNM$ includes numbers of $g_c$
coming from $m>0$ and $m<-1$.

\begin{figure}[h]
\includegraphics[scale=1.]{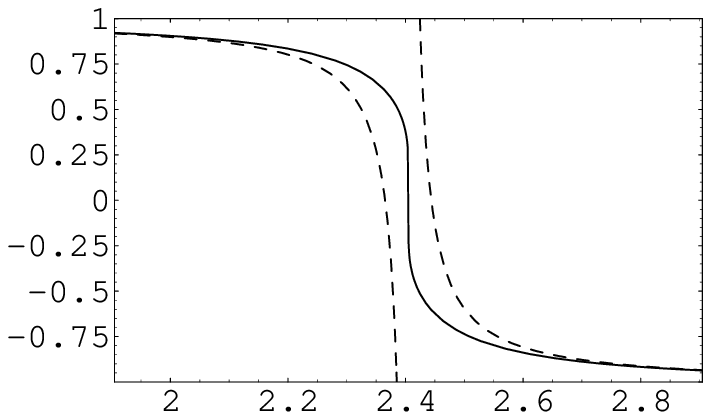}
\begin{picture}(0,0)(0,0)
\put(-102,-8){$g$}
 \put(-220,68){\rotatebox{90}{$F(g,r)$}}
 \end{picture}
\caption{Dependence of the  function  $F(g,r)$, defined in
Eq.(\ref{ChargeF}), on $g$ at $R/r=0.1$ in the vicinity of a minimal
value of $g_c$. Exact result obtained from Eq.(\ref{Charge})  is
shown as  a solid line, the asymtotics, obtained with the use of
Eq.(\ref{ChargeA}), as a dashed line.}\label{FF}
\end{figure}

In order to illustrate behavior of the induced charge in the
vicinity of some critical point $g_c$, it is convenient to represent
$Q_>(r)$ and $Q_<(r)$ at $r\gg R$ as follows:
\begin{eqnarray}\label{ChargeF}
Q_>(r)&=&eN[\mathrm{sign}(g-g_c)+F(g,r)]\, ,\nonumber\\
Q_<(r)&=&eN[\mathrm{sign}(g_c-g)-F(g,r)+M] \, , 
\end{eqnarray}
where $F(g,r)$ is some continuous  function of $g$. The dependence
of this function on $g$ at $R/r=0.1$, obtained from
Eq.(\ref{Charge}) in the vicinity of minimal value of $g_c$, is
shown in Fig.\ref{FF} (solid line), as well as its asymtotics,
obtained with the use of Eq.(\ref{ChargeA}) (dashed line).

It is seen that the region, where Eq.(\ref{ChargeA}) is not
applicable, is very narrow.

\section{Conclusion}\label{section7}
In this paper we have calculated the induced charge density
generated by the potential well in graphene at large distances.
Besides, we have obtained the induced charges outside of the radius
$r\gg R$ and inside of this radius for subcritical and supercritical
regimes.  Small variation of the potential parameters drastically
changes the induced charge density in the vicinity of the critical
values of $g$.

We are very grateful to O.P.~Sushkov, G.G.~Kirilin, and R.N.~Lee for
valuable discussions. The work was supported in part by  RFBR
 grants  08-02-91969 and 09-02-00024.

\end{document}